# ARCHITECTURE FOR REAL TIME CONTINUOUS SORTING ON LARGE WIDTH DATA VOLUME FOR FPGA BASED APPLICATIONS

Rourab Paul[1], Suman Sau[2a], Amlan Chakrabarti[2b]
Electronic Science Department[1]; A. K. Choudhury School of I.T [2], University of Calcutta, Kolkata-700009, India
Email: *rourabpaul@gmail.com*[1], *sumansau@gmail.com*[2a], *acakcs@caluniv.ac.in*[2b]

*Abstract* – **In engineering applications sorting is an important and widely studied problem where execution speed and resources used for computation are of extreme importance, especially if we think about real time data processing. Most of the traditional sorting techniques compute the process after receiving all of the data and hence the process needs large amount of resources for data storage. So, suitable design strategy needs to be adopted if we wish to sort a large amount of data in real time, which essential means higher speed of process execution and utilization of fewer resources in most of the cases. This paper proposes a single chip scalable architecture based on Field Programmable Gate Array (FPGA), for a modified counting sort algorithm where data acquisition and sorting is being done in real time scenario. Our design promises to work efficiently, where data can be accepted in the run time scenario without any need of prior storage of data and also the execution speed of our algorithm is invariant to the length of the data stream. The proposed design is implemented and verified on Spartan 3E(XC3S500E-FG320) FPGA system. The results prove that our design is better in terms of some of the design parameters compared to the existing research works.**

## I. INTRODUCTION

Sorting is required in many applications like image processing, video processing [1][2], ATM switching [3] where we may need to handle a large width of data. It is comparatively more critical in a real time scenario, where we generally have a long stream of information having a large data width and which needs to be sorted with a minimum amount of memory resources. We are proposing an idea of a continuous sorting process where we need not to store all of the information before the start of the sorting process i.e. we can go on executing the sorting process on the data which is coming in the run time. The memory management is handled by creating buffer arrays which can store the stream data in queue.

A kind of parallel sorting process is proposed by Devi Prasad et al. [4], where inputs are received by some parallel interfaces and after sorting process the sorted data are sent in a parallel way. Parallel computing gives the highest flexibility in this domain, which also results to an increase in processing speed but heat dissipation and area are increased which makes it difficult to implement the process in a constrained system with smaller resources, such as real time embedded systems. Sorting algorithm described in [3, 5] is not a choice in cost-sensitive system designs and consumes an enormous amount of FPGA resources. Our proposed architecture works on the less number of resources, which suits best for real time embedded system design. In the algorithm by Kumara Ratnayake et al. [6], the sorting time is invariant of the information size and this is based on [7]. The processing time of [6] only depends on the bit width of the data, not on the number of independent data to be sorted and hence we have adopted this algorithm with some modifications in our design. .

Our proposed work is an FPGA based design and implementation of the sorting algorithm, for real time data acquisition applications. In video processing [1] where large numbers of image frames are generated with time stamp values, we need such type of scalable real-time and compact sorting process over large volume and width of the data, so that the analysis can be done suitably. Here the volume of the data can be theoretically of infinite number of frames and large width means that each frame is of *n* number of bits, where n~48(test case), and may be more. The efficient buffer management in a real-time scenario is also a major contribution in our work.

In our design we have assumed that the data source is generating data frames having a fixed bit length and a portion of this data frame carries fixed length time stamp information as shown in Figure 1, which are generally generated from real events, and this type of scenario is quiet common in various real time applications [8]. The data acquisition system based on FPGA is a sort of data combiner here, which can accept data through its various data ports connected to various data channels of the actual data source [8]. At the other end of the FPGA is the backbone system, which will do the analysis of the data, and for which the time information of the data is of vital importance. The FPGA system here, is to do a real time sorting of the data frames coming through its various input channels based on their individual time stamps and it will send the sorted sequence of data frames to the backbone system so that the analysis can be done in offline.

The time stamp of the data is padded with information uniquely, for each of the recorded event with respect to a global clock. Sorting algorithm extracts the time stamp from the large width data packet and sorts the data packets according to their time stamp values. Validations of the proposed algorithm have been done on the Xilinx Spartan 3E (XC3S500E-FG320) FPGA board. The software code is written in System C and Xilinx Platform Studio (XPS) is used for the architectural design of the system.

The organization of the paper is as follows, Section 2 describes the existing algorithm [6] and modified algorithm for sorting, Section 3 detail out the proposed modification of the algorithm for real time implementation, Section 4 discusses on the target hardware architecture, Section 5 briefs on the results of the synthesis and implementation of the design and the concluding remarks are presented in Section 6.

## II. DATA SORTING ALGORITHM

Generally we have two options for designing the hardware of our proposed system
- Application specific Integrated Circuit (ASIC),
- Field Programmable Gate Array (FPGA).

FPGAs attracted notice, since their functionality can be changed (based on reconfiguration) although their throughput and performance is comparable to ASICs (based on concurrency and pipelining).

### A. Sorting Algorithm

We discuss the algorithm (*Algorithm 1*), which is a modification of [6]. We start with *N* number of data elements by first counting the number of occurrences for every element in the input unsorted array $A[0,....,N-1]$. The $i^{th}$ element of another array, $B[0,....,2^k-1]$ is storing how many times the $i^{th}$ number is present in $A[0,....,N-1]$. For example if we have the input array $A=[0,5,2,2,7,4]$ ( here $N=6$ and $k=3$) . In the unsorted array $A[i]$ number '0', '5', '4', '7' have no repetition, but the number '2' has a repetition, so the $0^{th}$, $5^{th}$, $4^{th}$, & $7^{th}$ element of array $B[i]$ have a rank of '1', & the $2^{nd}$ element has rank of '2'. Since the numbers '1', '3', & '6' are absent in the input array $A[i]$, $1^{st}$, $3^{th}$ & $6^{th}$ element of array $B[i]$ has rank of '0'. So in *Step 4* and *Step 5* of *Algorithm 1* after the loop iteration from 0 to N-1 the array $B[i]$ will look as $[1,0,2,0,1,1,0,1]$. In *Step 8* the rank of the each element is determined by counting the number of elements less than or equal to the element being considered, and is sorted in the array $C[0,....,2^k-1]$. Array $C[i]$ is responsible to place the elements of input array $A[i]$ into the sorted array $D[i]$ according to its amplitude. In our example $C=[1,1,3,3,4,5,5,6]$. Thus each element in the $A[i]$ is read and we store the value in the array $D[0,....,N-1]$ (sorted array) at the location stored in the array $C[0,....,2^k-1]$. If there are multiple occurrences for any element in $A[0,....,N-1]$, the next position is obtained by simply subtracting one from the corresponding location of $C[0,....,2^k-1]$, and the new result is stored back into $C[0,....,2^k-1]$. Once all the elements in the input array are read and scanned for the location of the array $D[1,....,N]$ from the array $C[0,....,2^k-1]$, the sorted result will be available in $D[0,....,N-1]$.

If we compare the existing algorithm [6] and our modified *Algorithm 1* we can find two modifications . First in *Step 12* of the algorithm [6] has C[A[i]] ← C[A[i]] +1 ,the modification in *Algorithm 1* is the decrement of C[A[i]] ← C[A[i]] -1. Secondly in *Step 13* of *Algorithm 1* [6] we return D [1 to *N*] instead of D [0 to *N-1*] [6], which does not add any additional cost to hardware resources and timing complexities. *Algorithm 1* is briefed as below:

**Algorithm 1**:

Input : A[0,....,N-1], N integer keys in the range of 0 to $2^k$-1 to be sorted.

Output : D[ 0,....,N-1]   Sorted keys

1.//initialize all the cells of the array B[ 0,....,$2^k$-1] to 0.
2.   B[ 0,...., $2^k$-1] ← 0
3.//Count same integers in A[ 0,....,N]
4. for i ← 0 to N-1 do
5.     B[A[i]] ← B[A[i]] + 1
6.// initialize C[0,....,$2^k$-1] and set C[0]=B[0]
7. for i ← 1 to $2^k$-1 do
8.     C[i] ← B[i] + C[i-1]
9.//Read each element from array    //A[0,....,N-1] starting from left   and //place them in D[0,....,N-1] at the //position given by  the content of C[ //0,....,$2^k$-1]
10. for i ← 0 to N-1 do
11.     D[C[A[i]]] ← A[i]
12.     C[A[i]] ← C[A[i]] -1
13. return D[ 1,....,N]

## III. HARDWARE BASED CONTINIOUS SORTING ALGORITHM

For implementing the above stated algorithm (*Algorithm 1*) with minimum hardware resources we will separate the data memory into two blocks. Each block has *N/2* elements and each of the elements will be stored using $2^k-1$ bits. A block (say $1^{st}$ block) of *N/2* width receives the input information from input interfaces and each element of the other block is (say $2^{nd}$ Block) initialized by zero. Then the two blocks are stored together into a buffer of N width where $1^{st}$, 0 to (N/2-1) set of data is occupied by the $1^{st}$ block and N/2 to (N-1) set of data are occupied by the $2^{nd}$ block. Now this buffer of N width is entered into the sorting process unit. Using the proposed algorithm the *N* numbers of information has been sorted out by the processor. The sorted result is stored in the array

*D[1,....,N]*. Now *D[ 1,...,N/2]* information has been sent as output to the hyper terminal of the computer for further process and *D[N/2+1,....,N]* information has been sent back to the 2nd block. In the next cycle 1st block receives the next *N/2* number of elements which adds up with the 2nd block elements and are again stored in the same buffer and this method is repeated in every consecutive cycle. The architecture of the process is shown in Figure 2. *Algorithm 2* is the modified hardware based version of *Algorithm 1* to extract the timestamp of 8 bits from the original information of 48 bits for a single frame of information. This is an arbitrary information structure where the width of the information, width of the time stamp and position of time stamp in information may be varied according to requirement.

**Algorithm 2**: Hardware based version of Algorithm 1

Input: X[0,….,3N-1], N integer keys in the range of 0 to $2^k$-1 to be sorted.
Output: D1[ 1,….,3N]   Sorted keys

1. //extract the time stamp
2. for i ← 0 to 3N-2  do
3.   X1[i+1] ←X[i+1]&0x00ff
4.   A[i/3] ←X1[i+1]
5.   i←i+2
6. //initialize array B[ 0,….,$2^k$-1] to 0.
7.  B[ 0,….,$2^k$-1] ←0
8. //Count same integers in A[ 0,….,N]
9. for i ← 0 to N-1 do
10.    B[A[i]] ← B[A[i]] + 1
11.// initialize C[ 0,….,$2^k$-1] and set       //C[0]=B[0]
12. for i ← 1 to $2^k$-1 do
13.    C[i] ← B[i] + C[i-1]
14.//Read each element from array //A[0,….,N-1] starting from left  and //place them in D[0,….,N-1] at the //position given by the content of C[ //0,….,$2^k$-1]
15. for i ← 0 to N do
16. D[C[A[i]]] ←X[((0x0003*i)+0x0001)];
17. D1[(0x0003*C[A[i]])-0x0001] ← D[C[A[i]]];
18. D1[(0x0003*C[A[i]])- 0x0002] ←X[(0x0003*i)];
19. D1[(0x0003*C[A[i]])] ←X[((0x0003*i)+0x0002)];
20.   C[A[i]] ←C[A[i]]-0x0001
21. return D[ 1,….,N]
_______________________________________

### IV. IMPLEMENTATION OF ARCHITECTURE

In real time data sorting scenario data needs to be sorted against their timing information that is padded with the data frames as time stamp information. Information which occurred first has least value of time stamp and the information that occurred last has the highest value of time stamp. The $2^k-1$ width of real time information is entered to the processor through some suitable interface and time stamp information works as the key information for the sorting process. Hence, before the sorting process starts we have to extract the time stamp from the information.  In our experiment we have taken a structure where information has 48 bits and the time stamp is padded from 16 bit to 23 bit of the information (see Figure 1).

We propose a further modification of the existing algorithm to fit it in the scenario of real data frames. Our 48 bit information is broken into 3 registers of 16-bit width (*X[0,....,(3N-1)]* (size of the array is *3N* because we are buffering 48 bits data into 3 register of 16 bits) and the time stamps are padded between 16 to 23 bit (this is arbitrary). In *Step 3* of the *Algorithm 2* we can see that we have masked the other bits except time stamp, and have taken all of the time stamps into the array *A[0,....,N-1]*. After sorting the array *A[0,....,N-1]* we passed the 48 bit information according to the time stamp to the array *D1[ 1,....,N]* and the *D[ 1,....,N]* is the completely sorted array based on the time stamp information. The detailed architecture is shown in Figure 4. One important thing that must be satisfied for faithful sorting is that, the lowest value of time stamps in $i^{th}$ frame must be greater than the highest value of timestamps in $(i-2)^{th}$ frame.

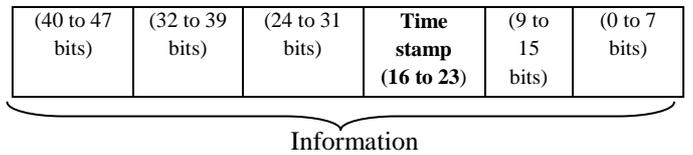

| (40 to 47 bits) | (32 to 39 bits) | (24 to 31 bits) | **Time stamp (16 to 23)** | (9 to 15 bits) | (0 to 7 bits) |

Information
Figure 1:  Information Structure

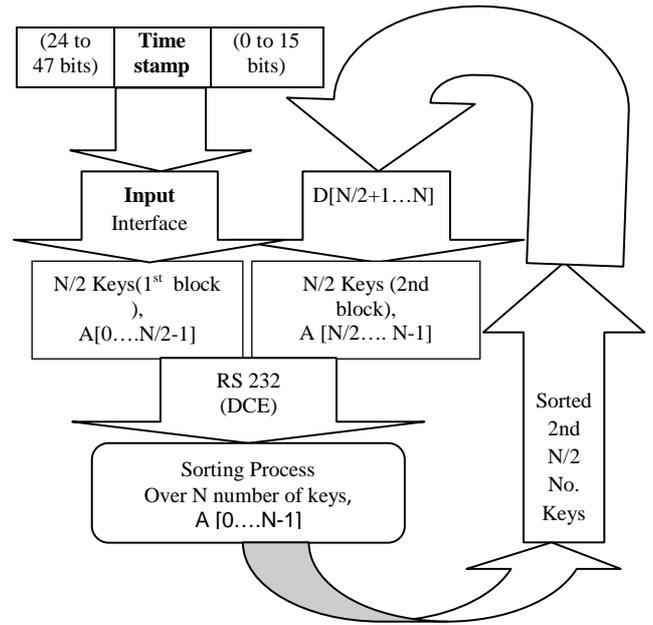

Figure 2: Proposed implementation flow diagram

TABLE1
POST SYNTHESIS UTILIZATION

| Resource Type | Used | Available | Percentage |
|---|---|---|---|
| Slices | 748 | 4656 | 16 |
| Slice Flip Flops | 917 | 9312 | 9 |
| 4 input LUT | 1420 | 9312 | 15 |

*A. Process Module of Proposed Algorithm*

The sorting process is done with the Xilinx Micro Blaze soft core processor with 50 MHz system clock. The processor uses Block RAM (BRAM) for storing instruction and intermediate data. The architecture is scalable and flexible since the implementation automatically computes the minimum required BRAMs.

*B. Real Time Data Interface:*

Data may be taken from real interfaces using serial port, PCI buses etc. into the FPGA board. In our experiment the input are given through the DIP switches and through the serial input using the keyboard interface. The result is sent to a computer through serial port and is displayed on Hyper Terminal of that machine. Speed of this serial communication can be controlled by the communication port setting (See Figure 5).

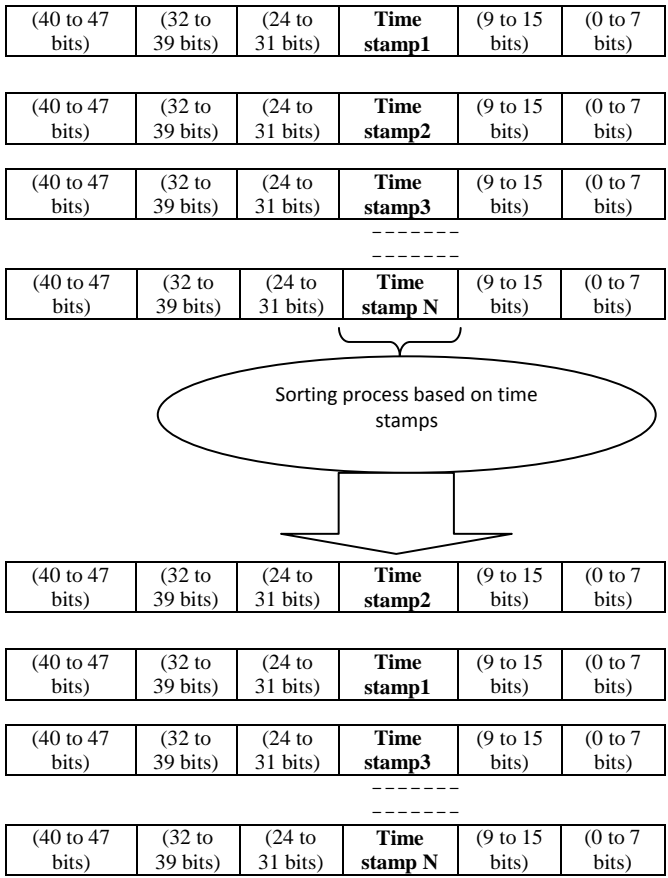

Fig3.Diagram of modified Algorithm based on time stamp

Figure 3 shows the modified algorithm architecture where *timestamp2<timestamp1< timestamp3 < timestamp N* (say).

V.SYNTHESIS AND IMPLEMENTATION RESULTS

The proposed architecture was synthesized using Xilinx ISE 11.1[9] and was implemented on XC3S500e Spartan 3E FPGA Board [9] with 20% Block Ram, 11% slice for 128 keys usage for 48 bit data and 8 bit time stamp. Table 1 shows the utilization of FPGA resources. The most significant part is that the whole architecture is a single-chip

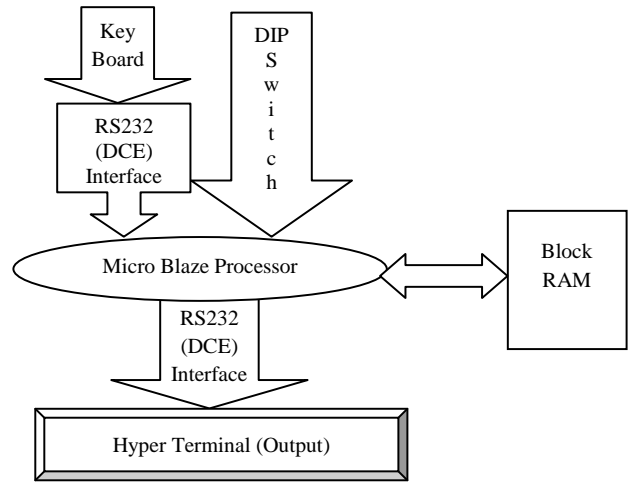

Figure 4: Detailed architecture using Xilinx FPGA

*A. Comparison with Existing Works*

The performance measure of our implementation is compared with existing methods as shown in Table 2. If we compare our results with [10] which uses a 0.35um CMOS based ASIC design, though the area utilization is much better in [10] (considering that Gate: Slice ratio is 10: 1 approx.) but our proposed FPGA architecture takes about 79% less clock cycle than [10] for the same clock frequency, also compared to [6] our design utilizes 28% less number of slices. This establishes the superiority of our design compared to the related research works. Fig 6 shows an output on Hyper terminal for 48 bit sorted information. So our design proposes an efficient solution for sorting a large number of information and using FPGA devices.

TABLE 2
COMPARISON TO THE EXIXSTING METHOD [10, 6]

|  | Existing[10] | Existing[6] | Proposed |
|---|---|---|---|
| Clock Frequency | 50 MHz | 133 MHz | 50 MHz |
| Clock cycle | 7936 | 128 | 2550 |
| Gates/Slices | 3185 Gates | 2784 Slices | 1665 Slices |
| Architecture | 0.35 µm CMOS | FPGA Vertex II | FPGA Spartan 3E |

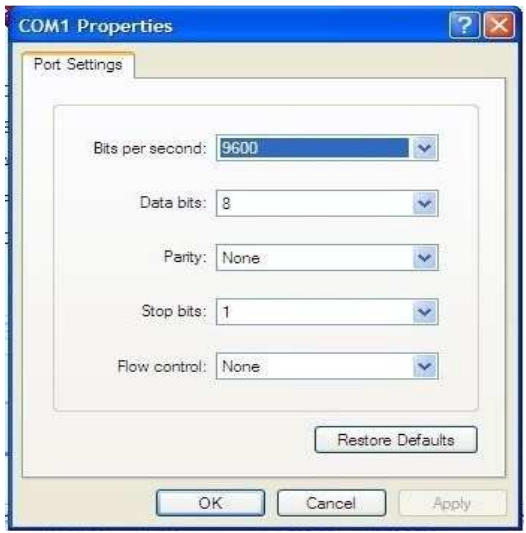

Figure 5: Configuration of the Hyper Terminal.

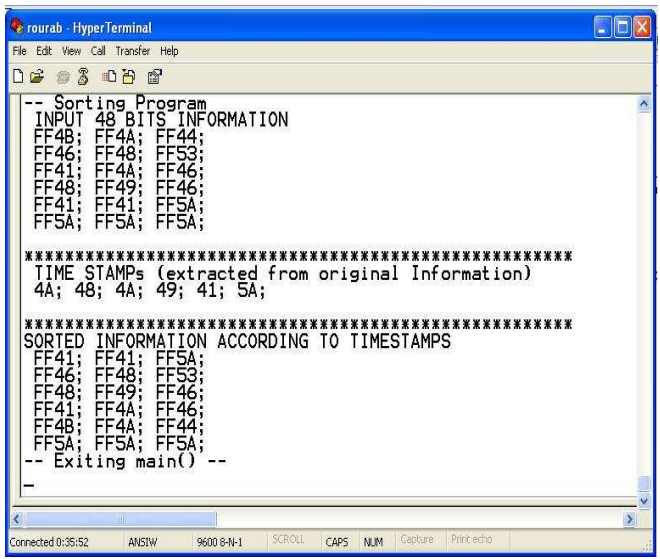

Figure 6: 48-bit output on hyper terminal

## VII. CONCLUSION

This paper proposes a single chip based architectural design and implementation of sorting large volume of data based on their time stamp values, which are received through its interfaces. The process does not store all of the data during the process to reduce the resource usage and efficient buffer management is devised to handle the need of real-time requirement. The proposed architecture is scalable to gain higher throughput by trading off only the FPGA internal memory resources (BRAM). The proposed implementation was successfully verified on an actual hardware platform that consists of a Xilinx Spartan 3E (XC3S500E) FPGA Board. The internal memory usage can be further minimized if the hardware platform with which the design is being implemented supports external SRAM. The DMA controller usage can also give a higher efficiency as it passes the information without interrupting the processor.

In future we would also attempt to use an RTOS environment ported on the FPGA platform, so that the sorting process can be called as a thread within a larger complex application process.

REFERENCES:

[1] A. Amer and E. Dubois, "Fast and reliable structure-oriented video noise estimation," IEEE Transactions on Circuits and Systems for Video Technology, vol. 15, no. 1, pp. 113-118, Jan 2005.

[2]Yan Lu, Ming Dai, Lei Jiang, Shi Li "Sort Optimization Algorithm of Median Filtering Based On FPGA"2010 International Conference of Machine Vision and Human Machine Interface.

[3] A.A. Colavita, A. Cicuttin, F. Frantik, and G. Capello, "Ortchip: A vlsi implementation of a hardware algorithm for continuous data sorting," IEEE Journal of Solid State Circuits, vol. 38, pp. 1076-1079, June 2003.

[4] Devi Prasad, Mohamad Yusri Mohamad Yusof, Smruti Santosh Palai, Ahmad Hafez Nawi Microelectronics Department, MIMOS Berhad. Technology Park Malaysia, Kuala Lumpur, Malaysia-57000. "Sorting networks on FPGA" Recent Researches in Telecommunications, Informatics, Electronics and Signal Processing. ISBN: 978-1-61804-005-3

[5] M. Bednara, 0. Beyer, J. Teich, and R. Wanka, "Tradeoff analysis and architecture design of a hybrid hardware/software sorter," IEEE International Conference on Application-Specific Systems, Architectures, and Processors, pp. 299-309, July 2000.

[6] Kumara Ratnayake and Aishy Amer " An FPGA Architecture of Stable-Sorting on a large data Volume: Application to Video Signal ".

[7] H. H Seward , "Information sorting in the application of electronic digital computer to business operation", M.S. Thesis, Massachusetts institute of technology (MIT), 1954.

[8]Design and Implementation of an Object Oriented Framework for Dynamic Partial Rec0nfiguration. Author(s): Abel, Norbert. https://www.gsi.de/documents/DOC-2011-May-41-1.pdf

[9] http://www.xilinx.com.

[10] C. Y. Huang, G. J. Yu, and B. D. Liu, "A hardware design approach for merge-sorting network," Proceedings of the 2001 IEEE International Conference on Circuits and Systems, vol. 4, pp. 534-537, May 2001